\documentclass[conference]{IEEEtran}
\IEEEoverridecommandlockouts

\usepackage{cite}
\usepackage{amsmath,amssymb,amsfonts}
\usepackage{graphicx}
\usepackage{booktabs}
\usepackage{multirow}
\usepackage{array}
\usepackage{url}
\usepackage{xcolor}
\usepackage{balance}
\def\BibTeX{{\rm B\kern-.05em{\sc i\kern-.025em b}\kern-.08em
    T\kern-.1667em\lower.7ex\hbox{E}\kern-.125emX}}
\begin{document}

\title{Inference as Flexibility: Ramp Management for Transmission-Connected AI Data Centres}

\author{\IEEEauthorblockN{Zhirui Liang}
\IEEEauthorblockA{
University of Michigan}}

\maketitle

\begin{abstract}
The rapid growth of large AI data centres introduces new operational challenges for power systems, including rapid ramping, oscillatory load behavior, voltage fluctuations, and supply-demand balancing impacts. For example, the Alberta Electric System Operator (AESO) has identified transmission-connected data centres (TCDCs) as large non-conforming loads that may need to limit their point-of-connection ramp rates. Existing mitigation approaches mainly rely on exogenous electrical resources, such as battery energy storage systems (BESS). This paper presents a proof-of-concept demonstration of a complementary software-defined mitigation layer: using flexible large language model (LLM) inference serving as endogenous TCDC flexibility to partially offset AI training power ramps. We consider a 150 MW TCDC with training, inference, and base-load components. A measured LLaMA-2-70B fine-tuning power profile is scaled to represent an aggregate training block, while measured LLaMA-3.1-70B inference power traces are used to model batch-size-dependent inference flexibility. Three strategies are compared: BESS-only mitigation, batch-size-only control, and coordinated batch-size plus BESS control. Simulation results show that the hybrid strategy reduces BESS discharge energy by 71\% and peak discharge power by 51\%, while maintaining near-complete compliance with a $\pm 10$ MW/min ramp limit.

\end{abstract}


\section{Introduction}

Large AI data centres are emerging as a new class of electricity demand in bulk power systems. Unlike conventional loads, they may operate many GPUs under synchronized training, fine-tuning, and inference workloads, creating rapid active-power changes. Recent studies show that large AI training jobs can exhibit significant power swings because compute-intensive phases consume substantially more power than communication-intensive phases~\cite{choukse2025power}. 

Such dynamics raise new challenges for transmission-connected data centres (TCDCs). As one example, the TCDC connection guide provided by Alberta Electric System Operator (AESO) identifies rapid ramping, oscillatory load behavior, voltage fluctuations, and supply-demand imbalance as key concerns for integrating large TCDCs~\cite{aeso2026guide}. The guide further identifies a 300 MW per 30 minute ramping capability, equivalent to 10 MW/min, and indicates that TCDCs may be required to limit ramping at the point of connection to this level during normal operation~\cite{aeso2026guide}. While this paper uses AESO's rule as a motivating example, the broader issue is how large AI data centres can shape their grid-facing ramps.

A natural mitigation strategy is to use fast-response electrical infrastructure, such as batteries, supercapacitors, and onsite generation~\cite{morovati2025bessdatacenter,ko2025mitigation}. These technologies provide valuable exogenous flexibility, but exclusive reliance on them can be costly because sufficient power and energy capacity must be installed, and frequent cycling can create operational costs through efficiency losses and battery degradation~\cite{he2020power}. This motivates the use of complementary internal flexibility resources.

This paper studies whether flexible LLM inference serving can assist ramp management for TCDCs. Since many AI data centres may co-locate training and inference workloads, inference flexibility can serve as a feasible \emph{endogenous} TCDC resource for compensating training-induced ramps. We assume a TCDC operator with control authority over both training and inference workload management.
Inference serving can be adjusted through parameters such as maximum batch-size configuration, admission rate, and routing; in particular, batch size affects inference power, token throughput, and inter-token latency~\cite{liang2026gputogrid,chung2026openg2g}.
Hence, inference flexibility is not free curtailment, but must be coordinated with computing-service requirements. In this paper, we present a proof-of-concept demonstration in which inference power is temporarily reduced during a training ramp-up, thereby reducing the residual ramp that must be compensated by the battery energy storage system (BESS). The contributions are: (i) framing LLM inference as a counter-ramping resource for TCDCs; (ii) proposing a coordination rule between inference batch size adjustment and BESS usage; and (iii) quantifying the effects on BESS power/energy, ramp violations, and inference throughput in a 150 MW TCDC case study.

\begin{figure}[t]
\centering
\includegraphics[width=\linewidth]{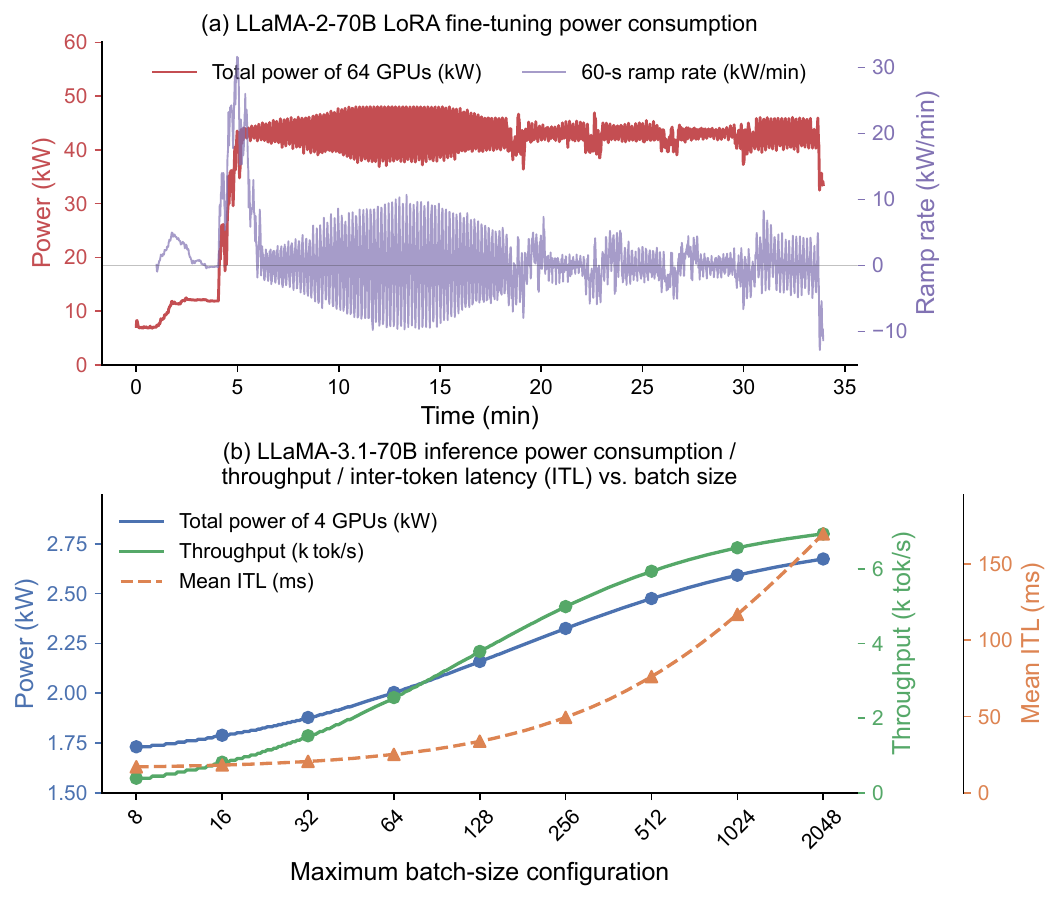}
\vspace{-2em}
\caption{Measured GPU workload characteristics: training power ramp and inference batch-size flexibility.}
\label{fig:training_inference_overview}
\end{figure}

\section{Simulation Setup}

We consider a 150 MW TCDC subject to a $\pm 10$ MW/min ramp envelope at the point of connection (POC), evaluated over a 60-second sliding window. The main setting, denoted \textbf{S0}, divides the facility load into 52.5 MW training load, 45 MW inference load, and 52.5 MW constant base load, corresponding to 35\% training, 30\% inference, and 35\% base load. The POC active power is
$P^{\mathrm{POC}}_t =
P^{\mathrm{base}} + P^{\mathrm{train}}_t + P^{\mathrm{infer}}_t(b_t) - P^{\mathrm{BESS}}_t$,
where positive $P^{\mathrm{BESS}}_t$ denotes BESS discharge, and $b_t$ is the inference batch size at time $t$. The POC ramp constraint is
$\left|P^{\mathrm{POC}}_t-P^{\mathrm{POC}}_{t-60\mathrm{s}}\right| \leq 10~\mathrm{MW}$.

The training profile is constructed from the NLR dataset, which reports per-GPU power measurements for LLaMA-2-70B LoRA fine-tuning on 16 nodes, or 64 NVIDIA H100 GPUs, across five repeated runs~\cite{vercellino2026measurement,nlr2026dataset}. Each GPU has a max thermal design power of up to 700~W~\cite{nvidia2026h100}. Fig.~\ref{fig:training_inference_overview}(a) shows the averaged power profile and its 60-second ramp rate. The trace contains a clear initialization period, a sharp start-up ramp, a quasi-steady training phase with oscillatory power variations, and a shutdown phase. These features illustrate why training and fine-tuning workloads can create grid-facing ramping challenges. We scale the averaged trace so that its peak equals the 52.5 MW training budget, treating the measured trace as a normalized power-shape template for an aggregate training block rather than a single facility-scale job. Without control, the scaled start-up and shutdown ramps exceed the $\pm 10$ MW/min POC envelope.

The inference fleet is represented by LLaMA-3.1-70B inference with tensor parallelism of 4 on H100 GPUs. For each maximum batch-size configuration, the measured power trace from the ML.ENERGY benchmark~\cite{mlenergy-neuripsdb25,mlenergy-github} is tiled over the simulation horizon and scaled to a 45 MW inference budget.
Fig.~\ref{fig:training_inference_overview}(b) shows the key tradeoff: increasing batch size raises inference power and token throughput, but also increases inter-token latency (ITL). Therefore, batch size cannot be made arbitrarily small or large. A small batch size sacrifices throughput, while a large batch size may violate latency requirements. In this case study, the ITL limit is set to 100 ms, so the maximum allowable batch size is 512. The available inference flexibility at any time is determined by the current batch size: reducing batch size provides downward power flexibility for mitigating training ramp-up, while later increasing batch size helps recover lost throughput.

We compare three rule-based strategies. In \textbf{Case 1}, inference runs at a fixed nominal batch size and BESS alone is used for ramp management. In \textbf{Case 2}, BESS is disabled and the inference batch size is adjusted within the feasible latency-constrained range to reduce POC ramp violations before gradually recovering to the nominal batch size. In \textbf{Case 3}, batch-size reduction is used first during training ramp-up events, while BESS compensates the residual ramp and supports batch-size recovery.

The controller observes the 60-second POC ramp and changes the batch size immediately when the ramp exceeds the $\pm 10$ MW/min envelope. To prevent chattering caused by short-term inter-batch power fluctuations, consecutive batch-size increases are subject to a 60-second minimum hold time. After the ramp remains within $\pm 5$ MW/min for 30 consecutive seconds, and at least 60 seconds have elapsed since the last recovery action, the controller moves the batch size one level back toward its nominal value. The nominal batch-size choice reflects the TCDC operator's service-quality priority. Keeping a lower nominal batch size preserves upward flexibility but reduces throughput, while operating near the latency-constrained maximum batch size improves throughput but leaves less upward headroom. These rules are not optimized; they are intended to demonstrate feasibility and quantify the potential BESS reduction enabled by inference flexibility.

\section{Simulation Results}

\begin{figure}[t]
\centering
\includegraphics[width=\linewidth]{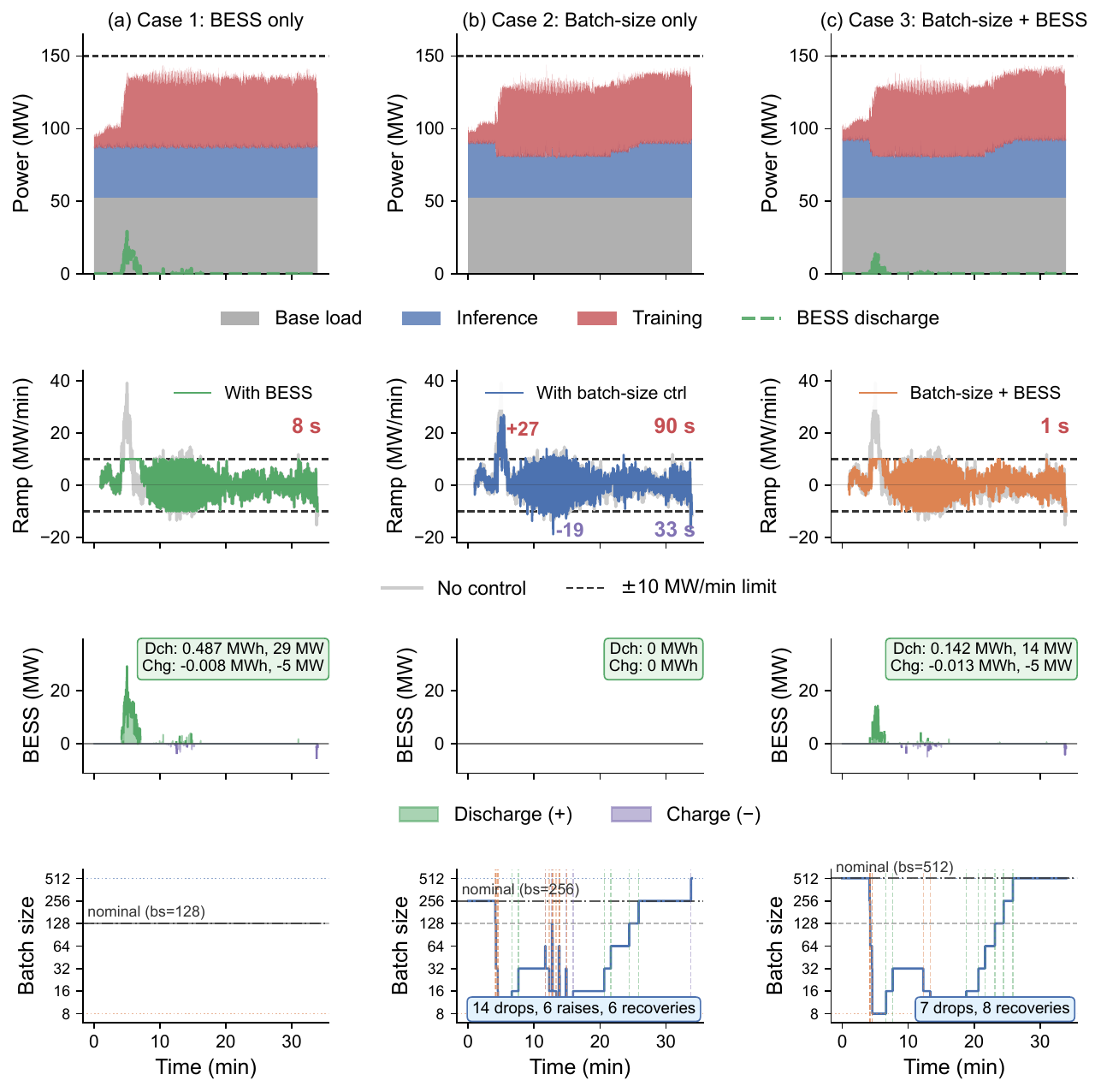}
\vspace{-2em}
\caption{Ramp-management results for the 150 MW TCDC under the S0 setting. Columns show Case 1 (BESS only), Case 2 (batch-size only), and Case 3 (batch size plus BESS). Rows report facility power decomposition, 60-second POC ramp rate, BESS charge/discharge power, and inference batch-size trajectory.}
\label{fig:main}
\end{figure}

Table~\ref{tab:main} summarizes the simulation results under three cases: BESS-only ramp management, batch-size-only ramp management, and the proposed hybrid strategy. In the S0 setting (35\% training, 30\% inference, and 35\% base load), the BESS-only case maintains near-complete ramp compliance, but requires 0.487 MWh of discharge energy and 29 MW peak discharge during the training start-up ramp. The batch-size-only case reduces inference power by dropping the batch size from 256 to 8 during start-up, but the available inference flexibility is insufficient to enforce the $\pm10$ MW/min limit by itself. The hybrid case first uses batch-size reduction to provide endogenous counter-ramping flexibility and then uses BESS for the residual ramp. This reduces BESS discharge energy by 71\% and peak discharge power by 51\%, respectively. The batch size later recovers to 512, limiting the loss of inference throughput. 
Fig.~\ref{fig:main} shows the detailed simulation results for the three cases under the S0 setting.

Table~\ref{tab:main} also reports two additional capacity splits: S1 with 40\% training and 25\% inference, and S2 with 30\% training and 35\% inference. A larger training fraction increases the residual ramp-management burden, as shown by the higher BESS discharge requirement in S1. A larger inference fraction increases the available endogenous flexibility, reducing both batch-only violations and hybrid BESS usage in S2. Across all settings, the hybrid strategy maintains near-complete ramp compliance while reducing BESS discharge energy by 52--86\% relative to the BESS-only case.

\begin{table}[t]
\centering
\caption{Simulation results under three capacity splits. S0: 35\% training / 30\% inference / 35\% base; S1: 40\% / 25\% / 35\%; S2: 30\% / 35\% / 35\%.}
\label{tab:main}
\footnotesize
\setlength{\tabcolsep}{3pt}
\begin{tabular}{llcccccc}
\toprule
Set. & Case & Throughput & ITL & $E_{\rm dch}$ & $E_{\rm chg}$ & $P_{\rm pk}$ & Ramp viol. \\
& & (M tok/s) & (ms) &  (MWh) &  (MWh) &  (MW) & up/down (s) \\
\midrule
S0 & BESS & 60.9 & 33.8 & 0.487 & 0.008 & 29 & 8/0 \\
S0 & Batch & 43.2 & 31.2 & 0 & 0 & 0 & 90/33 \\
S0 & Hybrid & 49.4 & 41.6 & 0.142 & 0.013 & 14 & 1/0 \\
\midrule
S1 & BESS & 50.8 & 33.8 & 0.705 & 0.021 & 33 & 10/0 \\
S1 & Batch & 29.4 & 28.1 & 0 & 0 & 0  & 128/48 \\
S1 & Hybrid & 38.7 & 40.2 & 0.337 & 0.025 & 21  & 12/0 \\
\midrule
S2 & BESS & 71.1 & 33.8 & 0.305 & 0.003 & 25 & 5/0 \\
S2 & Batch & 60.6 & 35.1 & 0 & 0 & 0 & 69/18 \\
S2 & Hybrid & 65.6 & 45.8 & 0.043 & 0.009 & 7 & 5/0 \\
\bottomrule
\multicolumn{8}{p{0.95\linewidth}}{\footnotesize Note: $E_{\rm dch}$ and $E_{\rm chg}$ are BESS discharge and charge energy, $P_{\rm pk}$ is peak BESS discharge power, and Ramp viol. reports up/down ramp-rate-violation duration.}
\end{tabular}
\end{table}

\section{Discussion and Conclusion}

The key insight is that inference flexibility and BESS provide complementary ramp-management capability. Batch-size control offers endogenous flexibility without additional hardware, but its power range is limited by throughput and latency requirements. BESS provides faster exogenous compensation, but requires additional power and energy capacity. The hybrid strategy exploits this complementarity by using inference batch-size reduction to absorb part of the training ramp-up and BESS to compensate the residual.

Ramp-down events reveal an important asymmetry. During training ramp-up, inference power can be reduced to offset the increasing training load. During training ramp-down, the controller would need to increase inference power, but this upward flexibility is limited by the maximum feasible batch size and latency constraint. Therefore, BESS remains valuable for sudden training shutdowns and batch-size recovery.

This paper is a proof-of-concept rather than an optimal control design. Future work should extend the control space beyond batch size, since GPU power caps, parallelism configurations, and dynamic voltage and frequency scaling (DVFS) may provide additional flexibility for grid-interactive LLM serving~\cite{hankendi2026pals}. Future work should also consider larger and more realistic TCDCs. Publicly compiled AESO-area data-center records include many proposed loads larger than the 150 MW case studied here~\cite{datacenterfyi2026aeso}. Such facilities may host multiple inference services and multiple training or fine-tuning jobs with different start times. Since different LLMs have different power ranges, latency constraints, and throughput values under batch-size control~\cite{chung2026openg2g}, selecting which workloads should provide flexibility becomes a co-optimization problem across workload scheduling, BESS dispatch, service quality, battery degradation, and grid-facing ramp compliance.

From a policy perspective, this proof-of-concept suggests that TCDC ramp-management requirements could be technology-neutral. Rather than prescribing hardware-only solutions, grid operators could allow TCDCs to demonstrate verified ramp-compliance capability using a portfolio of exogenous electrical resources and endogenous workload flexibility. Such flexibility should be represented through measurable operating envelopes that account for throughput, latency, and workload availability.

\balance
\bibliographystyle{ieeetr}
\bibliography{reference}

\end{document}